\documentclass[showpacs,amsmath,superscriptaddress,reprint,aps]{revtex4-2}

\usepackage{graphicx}
\usepackage{hyperref}
\usepackage{color}
\usepackage{float}
\usepackage{color}
\usepackage{lineno}

\begin{document}

\title
{Trion transfer in mixed-dimensional heterostructures}
\author{N.~Fang}
\email[Corresponding author. ]{nan.fang@riken.jp}
\affiliation{Nanoscale Quantum Photonics Laboratory, RIKEN Pioneering Research Institute, Saitama 351-0198, Japan}
\affiliation{Quantum Optoelectronics Research Team, RIKEN Center for Advanced Photonics, Saitama 351-0198, Japan}
\author{U.~Erkılıç}
\affiliation{Nanoscale Quantum Photonics Laboratory, RIKEN Pioneering Research Institute, Saitama 351-0198, Japan}
\affiliation{Quantum Optoelectronics Research Team, RIKEN Center for Advanced Photonics, Saitama 351-0198, Japan}
\author{Y.~R.~Chang}
\affiliation{Nanoscale Quantum Photonics Laboratory, RIKEN Pioneering Research Institute, Saitama 351-0198, Japan}
\affiliation{Quantum Optoelectronics Research Team, RIKEN Center for Advanced Photonics, Saitama 351-0198, Japan}
\author{S.~Fujii}
\affiliation{Quantum Optoelectronics Research Team, RIKEN Center for Advanced Photonics, Saitama 351-0198, Japan}
\affiliation{Department of Physics, Keio University, Yokohama 223-8522, Japan}
\author{D.~Yamashita}
\affiliation{Quantum Optoelectronics Research Team, RIKEN Center for Advanced Photonics, Saitama 351-0198, Japan}
\affiliation{Photonics-Electronics Integration Research Center, National Institute of Advanced Industrial Science and Technology (AIST), Ibaraki 305-8568, Japan}
\author{C.~F.~Fong}
\affiliation{Nanoscale Quantum Photonics Laboratory, RIKEN Pioneering Research Institute, Saitama 351-0198, Japan}
\affiliation{Quantum Optoelectronics Research Team, RIKEN Center for Advanced Photonics, Saitama 351-0198, Japan}
\author{S.~Morito}
\affiliation{Department of Chemistry, Saitama University, Saitama 338-8570, Japan}
\author{K.~Kanahashi}
\affiliation{Department of Materials Engineering, The University of Tokyo, Tokyo 113-8656, Japan}
\author{T.~Taniguchi}
\affiliation{Research Center for Materials Nanoarchitectonics (MANA), National Institute for Materials Science, Ibaraki 305-0044, Japan}
\author{K.~Watanabe}
\affiliation{Research Center for Electronic and Optical Materials, National Institute for Materials Science, Ibaraki 305-0044, Japan}
\author{K.~Ueno}
\affiliation{Department of Chemistry, Saitama University, Saitama 338-8570, Japan}
\author{K.~Nagashio}
\affiliation{Department of Materials Engineering, The University of Tokyo, Tokyo 113-8656, Japan}
\author{Y.~K.~Kato}
\email[Corresponding author. ]{yuichiro.kato@riken.jp}
\affiliation{Nanoscale Quantum Photonics Laboratory, RIKEN Pioneering Research Institute, Saitama 351-0198, Japan}
\affiliation{Quantum Optoelectronics Research Team, RIKEN Center for Advanced Photonics, Saitama 351-0198, Japan}

\begin{abstract}
Charged excitons, or trions, offering unique spin and charge degrees of freedom, have primarily been investigated in doped systems where charges are long considered indispensable. Here, we present an alternative route to ultra-efficient trion emission from an intrinsic, defect‑free semiconductor via a transfer mechanism. By exciting trions in two-dimensional tungsten-diselenide donors and transferring them into one-dimensional carbon-nanotube acceptors in mixed-dimensional heterostructures, we circumvent the usual carrier requirement, overcoming intrinsic Auger-quenching limitations. Benefitting from a reservoir effect induced by dimensional heterogeneity, this process achieves trion emission efficiencies increased by over 100-fold compared to conventional doping-based approaches, and remains robust across diverse doping conditions. Our findings extend the exciton-transfer paradigm to the three-body quasiparticles, offering a new platform for advancing excitonic physics and trion-based optoelectronic/spintronic applications.

\end{abstract}

\maketitle
\section{Introduction}
Excitonic processes lie at the heart of modern optics, photonics, and optoelectronics~\cite{High:2008,Paik:2019,Jauregui:2019,Kozlov:2019}. The transfer of excitons between dissimilar materials, where photoexcitation in one material leads to emission in another, is a fundamental phenomenon observed in organic semiconductors~\cite{Coles:2014NM}, quantum dots~\cite{Achermann:2004}, carbon nanotubes (CNTs)~\cite{Otsuka:2021,Fang:2023}, transition metal dichalcogenides (TMDCs)~\cite{kozawa:2016}, and other excitonic systems~\cite{Han:2020,Tabachnyk:2014}. Harnessing such exciton transfer can expand the absorption spectrum~\cite{Achermann:2004,kozawa:2016,Fang:2023}, boost quantum yields~\cite{Menke:2013}, and enable novel device concepts for light-emitting diodes~\cite{Ren:2021}, solar cells~\cite{Hardin:2009}, biosensing~\cite{Medintz:2005}, and other energy-harvesting applications~\cite{Cnops:2014}.

Another intriguing class of quasi-particles is the trion, a charged exciton formed by binding an exciton to an additional electron or hole. Trions carry net charge and spin, making them appealing for spintronics and quantum information technology~\cite{Schuetz:2010,Venanzi:2024,Mak:2013}. Since the early observations in 1993 within doped quantum wells~\cite{Kheng:1993}, trions have almost always been studied in doped materials~\cite{Mouri:2013NL,Liu:2019PRL,Anderson:2024}. Carriers are presumed essential for trion formation, which is typically supplied by electrostatic gating, chemical doping, or impurity doping. Besides the carriers that bind with excitons to form trions, the excess charges inevitably impact nearly every trion property. These charges limit trion emission through strong nonradiative Auger recombination~\cite{Javaux:2013,Lien:2019} and introduce many-body complexities that often necessitate describing trions within the exciton–polaron framework~\cite{Sidler:2017,Liu:2021}. Generating a “pure” trion flux in a charge- and trap‑free emitter, which would avoid the above interactions and enable potential trion-based spin qubits, has remained elusive under the existing paradigm.

Here, we introduce a fundamentally new concept—trion transfer—to achieve efficient trion emission from an intrinsically neutral, defect‑free emitter. It is realized in a mixed-dimensional heterostructure consisting of one-dimensional (1D) CNT and two-dimensional (2D) tungsten diselenide (WSe$_2$). Rather than doping the CNT, photoexcited trions in the 2D donor are transferred to the 1D acceptor, circumventing the usual nonradiative limitation imposed by free carriers. Through photoluminescence excitation, spatial imaging, and time-resolved measurements, we unveil a pronounced “trion reservoir'' effect arising from dimensional heterogeneity, achieving emission efficiencies exceeding those of conventional doping-based methods by over two orders of magnitude. In a field-effect transistor configuration where the CNT is tuned from neutral to highly doped state, the transfer-induced trion emission remains strong, distinguishing this free-carrier-insensitive mechanism from conventional optical processes. Moreover, by doping the WSe$_2$ donor with Nb, we show that trion transfer preserves high efficiency even in less-ordered systems where charge transfer coexists. These findings extend the fundamental understanding of the quasiparticle transfer process from excitons to trions, highlighting its unique optical route for achieving “pure” trion formation. A dense trion flux confined to a 1D channel opens new avenues in trion physics, spintronics, and advanced optoelectronic applications, for example the use of distinctive Fermi pressure absent in neutral excitons and the realization of trion superfluorescence.

\section{Results and discussion}
\paragraph*{Trion emission in mixed-dimensional heterostructures.}

\begin{figure*}
\includegraphics{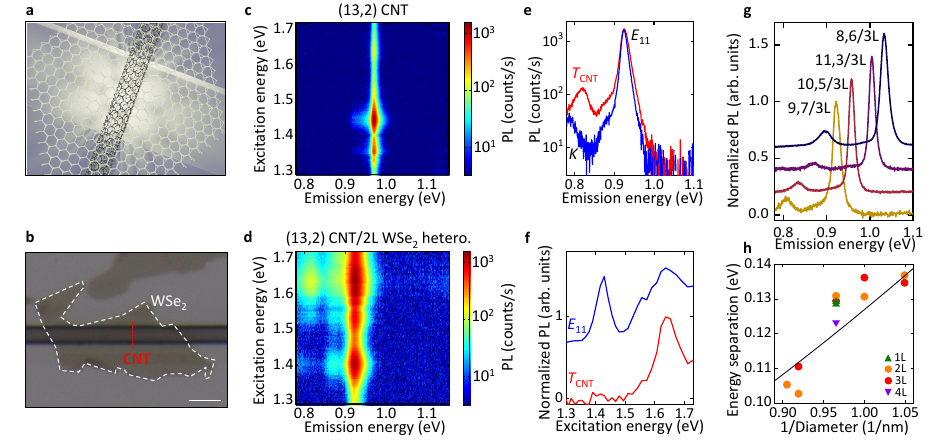}
\caption{
\label{Fig1} Universal trion emission in CNT/WSe$_2$ heterostructures.
\textbf{a} A schematic of a suspended CNT/WSe$_2$ heterostructure.
\textbf{b} A typical optical image of a CNT/2L WSe$_2$ sample. The CNT is indicated by the broken red line. The scale bar represents 3~$\mu$m.
\textbf{c} The PLE map of the pristine (13,2) CNT.
\textbf{d} The PLE map of the (13,2) CNT/2L WSe$_2$ heterostructure, where the tube differs from that in (\textbf{c}).
\textbf{e} The PL spectra of the (13,2) CNT/2L WSe$_2$ heterostructure in (\textbf{d}) at excitation under $E_{22}$ (blue, 1.420~eV) and $T_{\mathrm{WSe_2}}$/$X_{\mathrm{WSe_2}}$ (red, 1.634~eV), respectively.
\textbf{f} Normalized PLE spectra of integrated $E_{11}$ emission (blue) and $T_{\mathrm{CNT}}$ emission (red) in (\textbf{d}). The PL emission is integrated over a 10~meV-wide spectral window centered at the $E_{11}$ or $T_{\mathrm{CNT}}$ energies.
\textbf{g} Chirality-dependent PL spectra at $T_{\mathrm{WSe_2}}$/$X_{\mathrm{WSe_2}}$ excitation. We define a heterostructure nomenclature where (9,7) CNT/3L WSe$_2$ is represented by (9,7)/3L.
\textbf{h} Energy separation as a function of 1/Diameter from different samples. Green, orange, red, and purple symbols represent the heterostructures with 1L, 2L, 3L, and 4L WSe$_2$, respectively. The excitation power is 10~$\mu$W. The black line is a fit.
}
\end{figure*}

We construct the CNT/WSe$_2$ heterostructures by transferring a WSe$_2$ flake onto an individual air-suspended CNT using the anthracene-assisted technique~\cite{Otsuka:2021,Fang:2023}. The positions and chiralities of the CNTs are identified by photoluminescence (PL) spectroscopy~\cite{Ishii:2015} and the layer number of WSe$_2$ is determined before transfer, ensuring a clean and well-defined 1D/2D interface (Fig.~\ref{Fig1}a,b). Optical images confirm that all studied samples possess a large, uniform suspended WSe$_2$ region exceeding 5~$\mu$m in length that fully covers the CNTs. Notably, the CNTs in this study are defect-free and undoped as characterized by PL measurements~\cite{Ishii:2019}, whereas the natural WSe$_2$ crystals exhibit slight and unintentional doping due to intrinsic defects~\cite{Lyons:2019}.

Photoluminescence excitation (PLE) measurements provide direct insight into trion emission in these heterostructures. As a representative example, Fig.~\ref{Fig1}c shows a PLE map of a pristine (13,2) CNT, where the $E_{22}$ (1.447~eV) and $E_{11}$ (0.971~eV) transitions are clearly observed. In a (13,2)~CNT/bilayer (2L) WSe$_2$ heterostructure (Fig.~\ref{Fig1}d), these peaks redshift to 1.408~eV and 0.923~eV, respectively, due to the dielectric screening by WSe$_2$. The bright $E_{11}$ emission of the heterostructure suggests that the CNT remains largely undoped. Additionally, a new excitation peak at 1.643~eV arises from exciton transfer of the WSe$_2$ A excitons ($X_{\mathrm{WSe_2}}$) to the CNT $E_{11}$ excitons, denoted as $\lvert X_{\mathrm{WSe_2}}\rangle \rightarrow \lvert E_{11}\rangle$, consistent with transfer via tunneling for a type-I band alignment~\cite{Fang:2023}.

When excited near the $X_{\mathrm{WSe_2}}$ peak, a prominent low-energy emission peak emerges at 0.817~eV as shown in Fig.~\ref{Fig1}d. The energy separation $\Delta E$ between this subpeak and $E_{11}$ is 0.106~eV (Fig.~\ref{Fig1}e), which is smaller than the $\sim 0.140$~eV expected for CNT $K$-momentum excitons ($K$)~\cite{Kozawa:2024}. This suggests that the emission peak may originate from CNT trions ($T_{\mathrm{CNT}}$). The PLE spectrum in Fig.~\ref{Fig1}f indicates that the new emission peak is resonant with states near $X_{\mathrm{WSe_2}}$, rather than with $E_{22}$. In contrast, $E_{11}$ is resonant with both $X_{\mathrm{WSe_2}}$ and  $E_{22}$.

The low-energy emission peak is observed consistently in multiple samples. As shown in Fig.~\ref{Fig1}g, such peaks appear for (9,7), (10,5), (11,3), and (8,6)~CNT/3L~WSe$_2$ heterostructures. Moreover, similarly bright peaks also emerge in (10,5)~CNT heterostructures with WSe$_2$ layers ranging from 1L to 4L (Supplementary Note~1), indicating no clear layer number dependence within this range.

A defining characteristic of CNT trions is the diameter $d$ dependence of $\Delta E$~\cite{Tanaka:2019,Matsunaga:2011}, reflecting the $1/d$ scaling (Fig.~\ref{Fig1}h). The $\Delta E$ values observed here are close to those from surfactant-wrapped CNTs, suggesting a similar dielectric environment. $\Delta E$ is given by the equation:
\begin{equation}  \Delta E = \frac{A}{d} + \frac{B}{d^{2}},\, \end{equation} where A and B are constants for the binding energy and the singlet-triplet splitting, respectively. $A=60~\mathrm{meV}\cdot\mathrm{nm}$ and $B=67~\mathrm{meV}\cdot\mathrm{nm}^{2}$ reproduce the measured trend, which is comparable with surfactant-wrapped CNTs~\cite{Matsunaga:2011}; some deviations likely result from strain-induced shifts of $E_{11}$ during the formation of these suspended structures.

As with the exciton transfer process, the appearance of the trion peak also strongly correlates with the band alignment. Every sample exhibiting a pronounced $T_{\mathrm{CNT}}$ peak displays type-I band alignment, whereas the type-II heterostructures show neither $E_{11}$ nor $T_{\mathrm{CNT}}$ emission for excitation at the $X_{\mathrm{WSe_2}}$ energy (Supplementary Note~2).

\paragraph*{Trion transfer in mixed-dimensional heterostructures.}

\begin{figure*}
\includegraphics{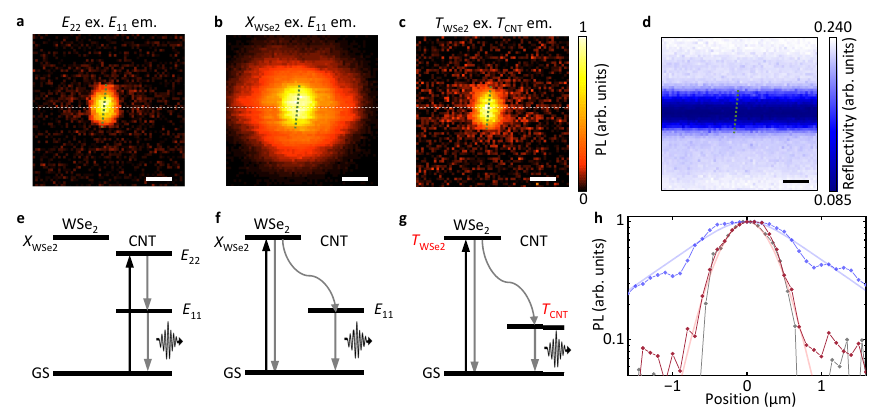}
\caption{\label{Fig2} PL excitation images for revealing trion transfer process.
\textbf{a--c} Normalized PL intensity maps from the (9,7) CNT/3L WSe$_2$ sample whose PL spectrum is shown in Fig.~\ref{Fig1}g. The PL images are constructed by integrating PL emission over a 30~meV-wide spectral window centered at $E_{11}$ energy (\textbf{a}, \textbf{b}) and $T_{\mathrm{CNT}}$ energy (\textbf{c}). The excitation is at $E_{22}$ (1.459~eV, \textbf{a}), $X_{\mathrm{WSe_2}}$ (1.653~eV, \textbf{b}), and $T_{\mathrm{WSe_2}}$ (1.653~eV, \textbf{c}). The excitation power is 10~$\mu$W. 
\textbf{d} The corresponding reflectivity image. The excitation is at $E_{22}$ (1.459~eV).
\textbf{e--g} Energy level diagrams showing the three processes occurring in \textbf{a--c}, respectively. GS indicates the ground state.
\textbf{h} Line profiles taken from \textbf{a--c}, as indicated by the white broken lines. The grey, blue, and red symbol-line plots are the experimental results from \textbf{a}, \textbf{b}, and \textbf{c}. The blue and red lines are corresponding fits. The scale bars in \textbf{a--d} represent 1~$\mu$m. The CNT is indicated by the broken green lines in \textbf{a--d}.
}
\end{figure*}

Our experiments suggest that $\lvert T_{\mathrm{CNT}}\rangle$ is generated via a trion transfer process. Specifically, we propose that photoexcited trions in WSe$_2$ lead to trion emission in CNTs, denoted by $\lvert T_{\mathrm{WSe_2}}\rangle \to \lvert T_{\mathrm{CNT}}\rangle$. This process differs from free-carrier-induced trion emission where any excitation that populates $\lvert E_{11}\rangle$ in a charged CNT can yield $\lvert T_{\mathrm{CNT}}\rangle$. Evidence supporting the absence of this free-carrier-based pathway includes the bright $E_{11}$ emission (largely unquenched by free carriers; Fig.~\ref{Fig1}c,d) and the lack of $T_{\mathrm{CNT}}$ under direct $E_{22}$ excitation (Fig.~\ref{Fig1}e,f). Instead, we propose that two transfer pathways coexist in the heterostructures: $\lvert X_{\mathrm{WSe_2}}\rangle \to \lvert E_{11}\rangle$ for excitons and $\lvert T_{\mathrm{WSe_2}}\rangle \to \lvert T_{\mathrm{CNT}}\rangle$ for trions. It is noted that $\lvert X_{\mathrm{WSe_2}}\rangle$ and $\lvert T_{\mathrm{WSe_2}}\rangle$ are nearly identical in energy (Fig.~\ref{Fig1}f) at room temperature owing to the small trion binding energy ($\sim 20$~meV)~\cite{Mak:2013}. However, the distinct spectral shapes (Fig.~\ref{Fig1}f) suggest that two processes arise from different initial states.

In fact, spatially resolved PL excitation imaging can distinguish the exciton and trion transfer processes. We perform the excitation imaging measurements on the (9,7)~CNT/3L WSe$_2$ sample using a three-dimensional motorized stage (Fig.~\ref{Fig2}a-c). The corresponding reflectivity image in Fig.~\ref{Fig2}d maps the trench over the same area. No other abrupt morphological changes are resolved, confirming that the characterized area, including the CNT, is uniformly covered by the WSe$_2$ flake. Under $E_{22}$ excitation, the $E_{11}$ emission profile matches the expected suspended CNT shape~\cite{Ishii:2015,Ishii:2019}, whereas excitation at the $X_{\mathrm{WSe_2}}/T_{\mathrm{WSe_2}}$ energy produces a broadened PL excitation image. This enlarged image indicates that WSe$_2$ A excitons excited at a distance funnel into the CNT after diffusion, acting as an exciton reservoir for exciting CNTs~\cite{Fang:2023}. Notably, under the same excitation energy, the $T_{\mathrm{CNT}}$ emission profile does not show such significant spatial broadening. Although $\lvert T_{\mathrm{WSe_2}}\rangle$ and $\lvert X_{\mathrm{WSe_2}}\rangle$ lie close in energy, the difference in the excitation images support a scenario in which $\lvert T_{\mathrm{WSe_2}}\rangle$ is initially excited in the trion transfer process and functions as a trion reservoir.

This difference in spatial broadening highlights distinct diffusion processes for $\lvert T_{\mathrm{WSe_2}}\rangle \to \lvert T_{\mathrm{CNT}}\rangle$ and $\lvert X_{\mathrm{WSe_2}}\rangle \to \lvert E_{11}\rangle$ (Fig.~\ref{Fig2}e-g). To quantify the diffusion lengths, we fit the PL line profiles in Fig.~\ref{Fig2}h with numerical solutions to the steady-state, one-dimensional diffusion equation:
\begin{equation} D \frac{d^2 n(x)}{dx^2} \;-\; \frac{n(x)}{\tau} \;+\; \frac{G}{\sqrt{2\pi r^2}} \exp\!\bigl(-\tfrac{2x^2}{r^2}\bigr) = 0, \end{equation} 
where $n$ is the trion (exciton) density that directly reflects the PL intensity, $x$ is the position along the trench, $D$ is the diffusion coefficient, $\tau$ is the lifetime, and $r = 0.58$~$\mu$m is the radius of the Gaussian laser profile. The best-fit simulation yields a diffusion length $L = \sqrt{D\,\tau} = 1.0$~$\mu$m for A excitons and only 0.15~$\mu$m for trions, in agreement with the slight trion spatial broadening beyond the laser spot. The results indicate that the trion diffusion length is approximately one order of magnitude smaller than that of the A exciton, which is consistent with previous studies of TMDCs reporting exciton and trion diffusion lengths of 1.5 and 0.3~$\mu$m, respectively~\cite{Uddin:2020}.

The relatively short trion diffusion length in WSe$_2$ limits $\lvert T_{\mathrm{WSe_2}}\rangle$ to a distance of $\sim 150$~nm around the CNT that may further contribute to the emission at $\lvert T_{\mathrm{CNT}}\rangle$ through transfer. It is however, still significantly more efficient than the direct excitation process, which is strictly confined to the $\sim 1$~nm CNT diameter. Moreover, we observe a pronounced trion reservoir effect in the time domain, as revealed by time-resolved PL (Supplementary Note~3). In a gate-doped suspended CNT under $E_{22}$ excitation, $\lvert T_{\mathrm{CNT}}\rangle$ decays with a 26~ps lifetime. In contrast, in the heterostructure under $T_{\mathrm{WSe_2}}$ excitation energy, it increases to 281~ps. Specifically, $\lvert T_{\mathrm{WSe_2}}\rangle$ with a lifetime of 521~ps continuously funnels into the CNT, acting as a trion reservoir in the time domain and extending the overall decay. Although transferred trions $T_{\mathrm{CNT}}$ in the heterostructure exhibits a shorter lifetime than $T_{\mathrm{WSe_2}}$ in pristine WSe$_2$, this does not indicate the emergence of nonradiative pathways. In fact, because CNTs are intrinsically undoped and defect-free, their bright excitons possess near-unity radiative quantum efficiency~\cite{Machiya:2022}. Trions transferred to the CNT bypass nonradiative loss channels in WSe$_2$ and could achieve similarly high quantum efficiency. By performing Monte Carlo simulations, we estimate that $\sim 20\%$ of trions excited in WSe$_2$ transfer to the CNT (Supplementary Note~4), which is a significant fraction given the dimensional mismatch here. Moreover, trion transfer time is estimated to be 1.3~ps, which is comparable with the fastest exciton transfer processes~\cite{Fang:2023,kozawa:2016,Qian:2008}.  

Trion transfer shares several features with exciton transfer, including comparable excitation energies and prolonged decay curves. Both processes exhibit linear excitation polarization independence (Supplementary Note~5), reflecting the two-dimensional nature of WSe$_2$ excitonic states. However, intercrossing processes between trions and excitons should not be allowed. Under the charge-conservation rule, transitions such as $\lvert T_{\mathrm{WSe_2}}\rangle \to \lvert E_{11}\rangle$ and $\lvert X_{\mathrm{WSe_2}}\rangle \to \lvert T_{\mathrm{CNT}}\rangle$ are not permitted. In addition, since bright excitons are spin-singlets and trions are spin-doublets, spin conservation also restricts transitions between quasi-particles of disparate spins. Consequently, no clear signature of population intermixing between excitons and trions is observed.

\paragraph*{Trion transfer overcoming free-charge-induced nonradiative limits.}

\begin{figure*}
\includegraphics{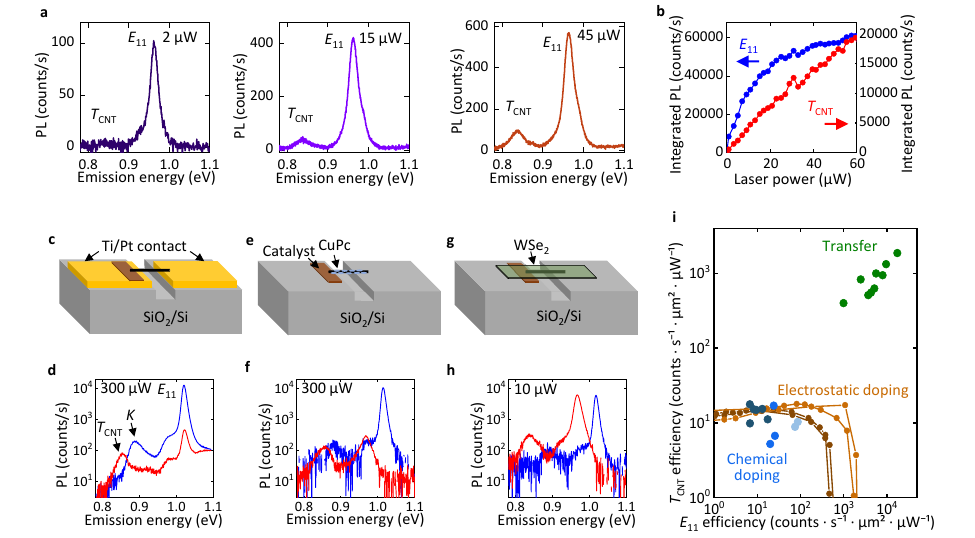}
\caption{\label{Fig3} Comparison between free-carrier induced trions and transfer-induced trions.
\textbf{a} PL spectra of the (10,5)~CNT/3L WSe$_2$ sample at powers of 2, 15, and 45~$\mu$W from left to right. The excitation is at $X_{\mathrm{WSe_2}}$/$T_{\mathrm{WSe_2}}$ energy.
\textbf{b} Integrated PL intensity as a function of the laser power for $E_{11}$ and $T_{\mathrm{CNT}}$, respectively. The PL emission is integrated over a 50~meV-wide spectral window centered at the $E_{11}$ or $T_{\mathrm{CNT}}$ energies.
\textbf{c} Schematic image of the suspended gated-CNT structure.
\textbf{d} PL spectra of the (10,5) gated CNT without (blue) and with $V_g = -0.8$~V (red). The excitation power is 300~$\mu$W and the excitation energy is $E_{22}$.
\textbf{e} Schematic image of the CNT/CuPc hybrid.
\textbf{f} PL spectra of the pristine (10,5) CNT (blue) and the (10,5) CNT/26-nm-thick CuPc hybrid (red). The excitation powers are 100~$\mu$W for the pristine CNT, 300~$\mu$W for the CNT/CuPc hybrid, and the excitation energy is $E_{22}$.
\textbf{g} Schematic image of the CNT/WSe$_2$ heterostructure.
\textbf{h} PL spectra of the (10,5) CNT before (blue) and after the formation of the heterostructure with a 1L WSe$_2$ flake (red). The excitation power is 10~$\mu$W and the excitation energy is $E_{22}$ for the pristine tube and $X_{\mathrm{WSe_2}}$/$T_{\mathrm{WSe_2}}$ for the heterostructure.
\textbf{i} $T_{\mathrm{CNT}}$ efficiency and $E_{11}$ efficiency from the three different structures. Green dots are from nine different CNT/WSe$_2$ heterostructures with an excitation power of 10~$\mu$W at $X_{\mathrm{WSe_2}}$/$T_{\mathrm{WSe_2}}$ energy. Brown symbol-line plots are from four suspended gated-CNT samples obtained by sweeping $V_g$ from 0~V to the positive side (light-brown two measured at 100~$\mu$W and dark-brown two at 300~$\mu$W). The excitation energy is $E_{22}$. Blue dots are from different CNT/CuPc samples. Light, medium, and dark blue indicate CuPc deposition thicknesses of 7, 16, and 26~nm, respectively. The excitation power is 300~$\mu$W and the excitation energy is $E_{22}$.
}
\end{figure*}

Unlike conventional trion formation which relies on free carriers in the emitter, trion transfer exploits the reservoir effect that promises brighter trion emission. To examine the increased emission efficiency, we measure the excitation power dependence in a (10,5)~CNT/3L WSe$_2$ sample (Fig.~\ref{Fig3}a). At a low power of 2~$\mu$W, $T_{\mathrm{CNT}}$ is much weaker than $E_{11}$. With increasing power, $T_{\mathrm{CNT}}$ increases more substantially than $E_{11}$. As shown in Fig.~\ref{Fig3}b, $E_{11}$ saturates at high powers due to exciton--exciton annihilation (EEA), whereas $T_{\mathrm{CNT}}$ increases linearly over the entire range. The results indicate that trion emission via transfer largely avoids the nonlinear inefficiency and suggest that trion--trion annihilation is minimal, possibly owing to electrostatic repulsion among trions.

To highlight the efficiency of transfer-based trion emission, we compare it with two standard trion-generation approaches: electrostatic doping and chemical doping (Fig.~\ref{Fig3}c--h). In the electrostatic method~\cite{Yoshida:2016}, we use a suspended (10,5)~CNT field-effect transistor and apply a back-gate voltage $V_g$ (Fig.~\ref{Fig3}c). Under $E_{22}$ excitation and zero gate bias, the PL spectrum exhibits a bright $E_{11}$ peak at 1.020~eV and a side peak $K$ at 0.887~eV. At $V_g = -0.8$~V, free carriers introduced into the CNT quench both $E_{11}$ and $K$, while $T_{\mathrm{CNT}}$ emerges at 0.856~eV (Fig.~\ref{Fig3}d). Because free-carrier-driven Auger recombination depletes excitons, the resulting trion emission remains weak and requires a high laser power of 300~$\mu$W to clearly resolve.

A similar phenomenon occurs with chemical doping. Depositing copper phthalocyanine (CuPc) onto suspended CNTs forms CNT/CuPc hybrids~\cite{Tanaka:2019}, introducing free carriers (Fig.~\ref{Fig3}e). Compared to a pristine (10,5)~CNT, the (10,5)~CNT/CuPc hybrid shows a strongly quenched $E_{11}$ and an emergence of $T_{\mathrm{CNT}}$ (Fig.~\ref{Fig3}f). Increasing the CuPc thickness from 7 to 26~nm systematically raises the doping level, thereby enhancing trion emission at the expense of $E_{11}$ emission (Fig.~\ref{Fig3}i). As with electrostatic doping, trion emission remains modest under the high power of 300~$\mu$W.

In contrast, transfer-induced trions exhibit a significantly stronger emission. Figure~\ref{Fig3}h shows PL spectra from a (10,5)~CNT before and after transferring a monolayer WSe$_2$ flake (Fig.~\ref{Fig3}g). Under direct $E_{22}$ excitation, the pristine CNT exhibits a clear $E_{11}$ peak. After forming the heterostructure and exciting at the $X_{\mathrm{WSe_2}}/T_{\mathrm{WSe_2}}$ energy, we observe bright $E_{11}$ and $T_{\mathrm{CNT}}$ peaks. The absence of free carriers in the CNT, combined with the reservoir effect for excitons and trions, yields the preserved $E_{11}$ emission and the significantly increased $T_{\mathrm{CNT}}$ emission at a low laser power of 10~$\mu$W.

To quantitatively compare these mechanisms, we define emission efficiencies for $T_{\mathrm{CNT}}$ and $E_{11}$ by normalizing each integrated PL intensity to the excitation power density (Fig.~\ref{Fig3}i) (see Methods and Supplementary Note 6). In four electrostatically doped CNTs, sweeping $V_g$ initially increases $T_{\mathrm{CNT}}$ efficiency, while $E_{11}$ efficiency is inevitably quenched. By further increasing $V_g$, $T_{\mathrm{CNT}}$ efficiency eventually saturates and then declines. Such saturation behavior with respect to carrier density is widely observed in trion-emitting systems~\cite{Ross:2013,Baek:2021,Ziegler:2023}, suggesting an intrinsic efficiency limit. Notably, at zero $V_g$ condition, $E_{11}$ efficiency is lower than in CNT/WSe$_2$ heterostructures. This difference arises from efficient exciton transfer in the heterostructures and strong EEA under high laser power in gated CNTs; the latter can be largely avoided by reducing the excitation from 300~$\mu$W to 100~$\mu$W. Meanwhile, at high $V_g$ condition where $E_{11}$ efficiency is severely degraded, $T_{\mathrm{CNT}}$ efficiency is insensitive to laser power, consistent with the absence of trion--trion annihilation. CNT/CuPc samples follow a similar trend with carrier density, which is modulated by CuPc thickness. We also note that the results are more scattered due to strain and morphological variations.

Transfer-based trion emission in all CNT/WSe$_2$ samples allows $E_{11}$ and $T_{\mathrm{CNT}}$ to coexist and scale together. Remarkably, the best samples achieve trion emission efficiencies more than two orders of magnitude above the limits of doping-based methods. Given that exciton transfer proceeds via a direct tunneling mechanism, it is plausible that trion transfer similarly involves simultaneous tunneling of an additional charge (Supplementary Note 2), and an efficient tunneling pathway enhances both processes. After trion emission, the extra charge likely tunnels back to WSe$_2$ under equilibrium conditions and the CNT remains undoped, as supported by the lack of any photodoping signature. Moreover, because the reservoir effect for both excitons and trions depends strongly on diffusion, WSe$_2$ flakes with fewer scattering centers should enhance both emissions, explaining the observed correlation in the generation of $\lvert E_{11}\rangle$ and $\lvert T_{\mathrm{CNT}}\rangle$. Improvements of tunneling efficiency and WSe$_2$ uniformity are thus expected to further increase trion emission, marking a conceptual difference from conventional doping-based approaches.

\paragraph*{Trion transfer in free-charge-modulated mixed-dimensional heterostructures.}

\begin{figure*}
\includegraphics{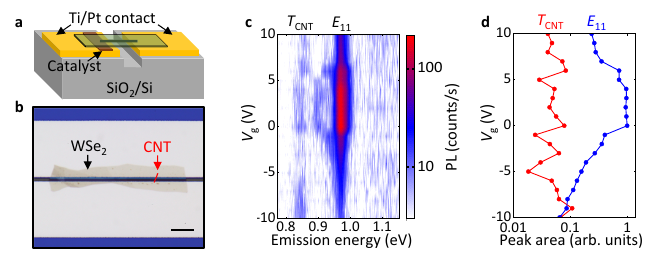}
\caption{\label{Fig4} Trion transfer in gated CNT/WSe$_2$ heterostructures.
\textbf{a} A schematic of a suspended gated CNT/WSe$_2$ heterostructure.
\textbf{b} An optical image of the (10,5) CNT/3L WSe$_2$ device. The scale bar represents 10~$\mu$m.
\textbf{c} PL spectra as a function of gate voltage. The excitation power is 2~$\mu$W and the excitation energy is $X_{\mathrm{WSe_2}}$/$T_{\mathrm{WSe_2}}$ at 1.642~eV.
\textbf{d} PL peak area for $T_{\mathrm{CNT}}$ (red) and $E_{11}$ (blue) in \textbf{c} as a function of gate voltage. The peak area is obtained by Lorentzian fitting at each $V_g$.}
\end{figure*}

Overcoming free-charge-induced limits suggests that free charges do not play a distinct role in the trion transfer process. To gain a comprehensive understanding of this mechanism, we deliberately introduce free carriers into the mixed-dimensional heterostructures via either field-effect gating or substitutional doping. We first achieve controlled doping in the nanotubes by fabricating a suspended (10,5)~CNT field-effect transistor and transferring a 3L~WSe$_2$ flake onto it (Fig.~\ref{Fig4}a,b). As natural WSe$_2$ crystals contain defects with a density of $10^{12}$~cm$^{-2}$~\cite{Tan:2025}, defect-induced gap states weaken the gate bias modulation effect at the suspended WSe$_2$ Fermi level. In contrast, the CNTs are characterized to be free of defects so that gate bias would effectively modulate the suspended CNT Fermi level~\cite{Yoshida:2016}.

Figure~\ref{Fig4}c shows the gate-dependent PL spectra of this heterostructure under $X_{\mathrm{WSe_2}}/T_{\mathrm{WSe_2}}$ excitation. $E_{11}$ is strongly quenched at high gate bias, indicating an effective field-effect modulation of the nanotube. We note that $E_{11}$ in the transistor configuration is slightly quenched at a gate bias of 0~V compared with the pristine CNT/WSe$_2$ heterostructures, likely due to additional charging of the CNT through the metal in contact with the WSe$_2$. Consequently, the charge-neutral point shifts from 0~V to $V_g = 2.9$~V. Remarkably, $T_{\mathrm{CNT}}$ emerges at the charge-neutral point, again underscoring the key feature of transfer-induced trions: they do not require free carriers in the CNT.

Such a “pure” trion flux free of extra carriers is highly relevant for trion-based device applications, which rely on manipulating spin and charge of trion qubits. Unlike free carriers, the excitons co-generated do not share the same spin or charge as trions, and can be further excluded by exploiting selection rules in the transfer process. By either lowering the operating temperature of the heterostructure transistor or choosing a donor material where exciton and trion energies are more distinctly separated, selective trion excitation can be achieved that leads to an even purer trion flux.

We also observe that trion emission remains largely insensitive to $V_g$ over the entire range, as clearly indicated by the $T_{\mathrm{CNT}}$ peak area versus $V_g$ in Fig.~\ref{Fig4}d. In another sample, $E_{11}$ is quenched by $\sim94\%$ at large negative $V_g$, yet $T_{\mathrm{CNT}}$ remains unquenched (Supplementary Note~7). Such a charge-density robustness distinguishes trion transfer not only from charge-induced trion emission but also from most other optical processes, where high carrier densities typically quench PL.

\begin{figure*}
\includegraphics{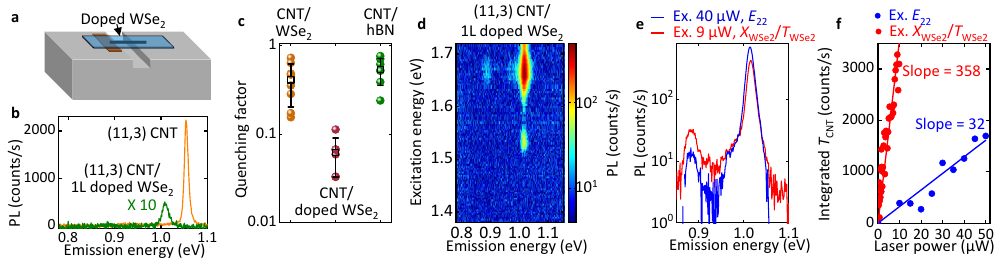}
\caption{\label{Fig5} Trion transfer in CNT/doped WSe$_2$ heterostructures.
\textbf{a} A schematic of a suspended CNT/WSe$_2$ heterostructure.
\textbf{b} PL spectra of the (11,3) CNT before (orange) and after the formation of the heterostructure with a 1L doped WSe$_2$ flake (green). The excitation power is 10~$\mu$W and the excitation energy is $E_{22}$.
\textbf{c} Quenching factor $Q$ for three structures. Error bars are the standard error of the mean.
\textbf{d} The PLE map of the (11,3)/1L doped WSe$_2$ heterostructure. The excitation power is 10~$\mu$W.
\textbf{e} The PL spectra of the (11,3)/1L doped WSe$_2$ heterostructure at $E_{22}$ (1.540~eV, 40~$\mu$W, blue) and $X_{\mathrm{WSe_2}}$/$T_{\mathrm{WSe_2}}$ (1.634~eV, 9~$\mu$W, red), respectively.
\textbf{f} Integrated PL intensity of $T_{\mathrm{CNT}}$ as a function of the laser power at $E_{22}$ (blue) and $X_{\mathrm{WSe_2}}$/$T_{\mathrm{WSe_2}}$ (red) from the (11,3)/1L doped WSe$_2$ heterostructure. The $T_{\mathrm{CNT}}$ emission is integrated over a 50~meV-wide spectral window centered at the peak energy.}
\end{figure*}

After examining the free-carrier effect on the CNT acceptor in the trion transfer process, we next focus on the WSe$_2$ donor by introducing Nb substitutional doping (Fig.~\ref{Fig5}a; see Methods). Incorporating Nb at $\sim8.8\times10^{18}$~cm$^{-3}$ produces a nondegenerate p-doped WSe$_2$ flake~\cite{Kanahashi:2025}. When the 1L flake is transferred onto a (11,3)~CNT, the $E_{11}$ peak redshifts by 0.045~eV, and the emission is quenched by more than 90\% (Fig.~\ref{Fig5}b). The doping in WSe$_2$ shifts its Fermi level, and charges move across the interface to align the Fermi levels when forming the heterostructure. Such ground-state charge transfer dopes the CNT and thereby quenches its PL.

To quantify this quenching, we define a quenching factor $Q$ as the $E_{11}$ PL intensity under $E_{22}$ excitation with respect to the same CNT prior to heterostructure formation. Figure~\ref{Fig5}c compares $Q$ for three heterostructures: CNT/WSe$_2$, CNT/doped~WSe$_2$, and CNT/hexagonal boron nitride (hBN). Because bulk hBN serves as an excellent insulator for CNTs~\cite{Fang:2020}, the quenching in CNT/hBN mainly reflects other factors such as strain from the transfer process rather than charge transfer. In CNT/WSe$_2$, many samples show similar high $Q$ values as in CNT/hBN, again confirming negligible free charges in the CNT. A few samples exhibit lower $Q$, likely due to significant strain from flexible thin WSe$_2$ flakes. In contrast, CNT/doped WSe$_2$ samples yield an average $Q$ of only 0.062, confirming strong quenching by free carriers through the charge transfer process.

Despite this charge-induced quenching, both exciton and trion transfer persist in CNT/doped WSe$_2$ heterostructures. As indicated in the PLE map in Fig.~\ref{Fig5}d, $E_{11}$ emission at 1.018~eV and $T_{\mathrm{CNT}}$ emission at 0.883~eV are resonantly excited at the $X_{\mathrm{WSe_2}}/T_{\mathrm{WSe_2}}$ energy of 1.670~eV. Since the CNT is now charged, we also expect trion emission under direct $E_{22}$ excitation. Indeed, a trion peak appears under $E_{22}$ excitation energies (Fig.~\ref{Fig5}e), confirming charge transfer. However, even at a higher laser power of 40~$\mu$W, $T_{\mathrm{CNT}}$ emission via $E_{22}$ remains much weaker than via trion transfer at 9~$\mu$W, consistent with the higher efficiency of the latter (Fig.~\ref{Fig5}f). From the power dependence in Fig.~\ref{Fig5}f, the trion peak increases linearly with laser power, showing a slope of 358~counts$\cdot$s$^{-1}\cdot\mu$W$^{-1}$ under $X_{\mathrm{WSe_2}}/T_{\mathrm{WSe_2}}$ excitation, consistent with the behavior in CNT/WSe$_2$ samples (Fig.~\ref{Fig3}a). In contrast, direct $E_{22}$ excitation yields a slope of only 32~counts$\cdot$s$^{-1}\cdot\mu$W$^{-1}$, over an order of magnitude smaller.

Although transfer-induced trion emission in the CNT/doped~WSe$_2$ sample far exceeds that of free-carrier-based trion generation methods, it is weaker than in the undoped case (Fig.~\ref{Fig3}h and Fig.~\ref{Fig5}e). Note that doping the CNT itself does not necessarily degrade trion transfer (Fig.~\ref{Fig4}). Instead, we attribute the reduced trion emission to the impact of Nb dopants on the spatial reservoir effect for trions in WSe$_2$. PL excitation images (Supplementary Note~8) show much reduced diffusion, with the exciton diffusion length decreasing to 0.12~$\mu$m and the trion length unresolvable. Because trions carry net charge and have relatively small binding energies, they are expected to be more susceptible to scattering, likely restricting their diffusion to 10--100~nm. Consequently, the diminished reservoir effect can attenuate trion emission compared to undoped WSe$_2$.

The coexistence of charge, exciton, and trion transfer in the doped 1D/2D heterostructures offers broader insights for other material systems. We anticipate that a variety of low‑loss 1D or 0D emitters, such as nanotubes, nanowires, or quantum dots, could likewise accept trions from higher‑dimensional donors efficiently without suffering strong nonradiative recombinations. It also suggests that in heterostructures such as 2D/2D assemblies where no dimensional heterogeneity exists, trion reservoir effect may be weaker. However, the key hallmarks of trion transfer, resonance with the excitation states in the donor and insensitivity to the free carriers in the acceptor, should still hold, which provides a versatile route to trion generation and manipulation.

A reservoir-fed high-density trion population confined to a 1D channel raises the prospect of superfluorescence, where the cooperative emission scales as the square of the participating population. Trion-mediated optical gain can couple readily to small-mode-volume, high-Q cavities for cavity quantum electrodynamical effects and low-threshold lasing. With a 1D trion flux, strong trion–trion repulsion is anticipated, which maybe manifested in the negligible trion–trion annihilation in Fig.~\ref{Fig3}b. As charged composite fermions, trions experience long-range Coulomb repulsion and an effective Fermi pressure that exceed the short-range exchange and phase-space-filling interactions governing neutral excitons. Under these Fermi-pressure–driven constraints, a trion flux may support long-range coherent transport and enable optoelectronic functionalities with both electrical and optical readout. Moreover, their internal spin configurations could serve as qubit degrees of freedom in quantum computing, with further effort on robust initialization, coherent control, and high-fidelity readout.

In conclusion, we have demonstrated a conceptually new optical mechanism, trion transfer, in mixed-dimensional heterostructures formed by 1D CNTs and 2D WSe$_2$. By circumventing carrier doping in the CNT, this process overcomes the typical nonradiative Auger limitations and yields trion emission efficiencies exceeding conventional methods by more than two orders of magnitude. Detailed measurements, including photoluminescence excitation, spatial mapping, and time-resolved PL, reveal how trions photoexcited in WSe$_2$ can diffuse and transfer into the CNT without introducing free carriers. This mechanism not only preserves bright exciton emission but also remains robust under CNT gating or WSe$_2$ doping. Our findings establish that dimensional heterogeneity---a 1D acceptor combined with a 2D donor---provides a unique trion reservoir effect that can be harnessed for efficient trion generation. Extending these concepts to other low-dimensional systems could open new opportunities in many-body excitonic physics, spin/valleytronics, and advanced optoelectronics, paving the way for novel trion-based devices.

\section*{Methods}
\paragraph*{Air-suspended carbon nanotube growth.}
To prepare air-suspended CNTs, we start from silicon dioxide (SiO$_2$)/silicon (Si) substrates with pre-fabricated trenches~\cite{Ishii:2015,Ishii:2019}. Electron-beam lithography (EBL)  is used to pattern alignment markers and trenches on the Si substrate, followed by dry etching to form trenches approximately 900~$\mu$m in length and 0.5--2.0~$\mu$m in width. A thermal oxidation step then grows a 60--70~nm SiO$_2$ film inside the trenches. Next, another round of EBL defines the areas where an iron (Fe) film ($\sim1.5$~\AA\ thick) is deposited by electron-beam evaporation as a catalyst for CNT growth. Finally, CNTs are synthesized by alcohol chemical vapor deposition at 800~$^\circ$C for 1~minute. By optimizing the Fe thickness, we predominantly obtain isolated and high-quality CNTs. For heterostructure fabrication with WSe$_2$, we select isolated, defect-free, chirality-identified CNTs with lengths of 1.0--2.0~$\mu$m. Specifically, the tubes are confirmed to exhibit bright PL emission with a small linewidth of $\sim10$~meV, a smooth suspended single-tube shape in PL imaging without any trapping or quenching sites, and a high linear polarization degree over 90\%.

\paragraph*{Anthracene crystal growth.}
Anthracene crystals used for stamping WSe$_2$ flakes onto CNTs are grown by in-air sublimation~\cite{Otsuka:2021,Fang:2023}. Anthracene powder is placed on a glass slide maintained at 80~$^\circ$C. A second glass slide, typically $\sim1$~mm above the anthracene source, collects the sublimated material. Thin, large-area single crystals grow on the upper slide over 10~hours. To promote the formation of thin, large-area crystals, ink patterns from a commercial marker are applied to the glass, inhibiting 3D crystal nucleation.

\paragraph*{2D materials preparation.}
Natural WSe$_2$ crystals are purchased from HQ Graphene, and their quality is characterized by electrical measurement~\cite{Fang:2018}. They exhibit a relatively high defect density of $\sim 10^{12}$~cm$^{-2}$ but remain nearly neutral with only slight carrier doping because most defects are deep-level states. Nb-doped WSe$_2$ crystals are synthesized through chemical vapor transport~\cite{Kanahashi:2025}. In this growth, powders of W, Nb, and Se are uniformly distributed within a quartz ampoule and placed into a tube furnace. Nb-doped WSe$_2$ single crystals grow across the ampoule.

\paragraph*{WSe$_2$ transfer using anthracene crystals.}
WSe$_2$ flakes are mechanically exfoliated onto standard 90~nm SiO$_2$/Si substrates. Their layer thickness is determined by optical contrast. A polydimethylsiloxane (PDMS) stamp supported on a glass slide is used to pick up a single anthracene crystal, forming an anthracene/PDMS stamp. The anthracene crystal then collects the targeted WSe$_2$ flake from the substrate by pressing the stamp onto the flake and withdrawing it rapidly ($>10$~mm/s). Next, the same stamp is aligned over a chirality-identified CNT, whose location is determined via previous measurements, and peeled off slowly ($<0.2$~$\mu$m/s) so the anthracene/WSe$_2$ stack adheres to the receiving substrate. The anthracene crystal is then removed by sublimation in air at 110~$^\circ$C for $\sim10$~minutes, leaving behind a clean, fully suspended CNT/WSe$_2$ heterostructure. This dry transfer protocol minimizes contamination, and the anthracene crystal mechanically protects both the CNT and the WSe$_2$ flake during transfer~\cite{Otsuka:2021,Fang:2023}.

\paragraph*{CNT/hBN heterostructure preparation.}
The thick hBN flakes ranging from 20 to 90~nm are first prepared on PDMS by mechanical exfoliation from crystals supplied by NIMS, and then transferred onto the target CNTs at 120~$^\circ$C using a micromanipulator system~\cite{Fang:2020}.

\paragraph*{CuPc deposition on CNTs.}
The chip with CNTs is placed in a vacuum chamber for the evaporation of CuPc (Sigma-Aldrich) and is maintained at about 80~$^\circ$C for 10~minutes to remove adsorbed molecules on the CNTs~\cite{Tanaka:2019}. CuPc molecules are deposited on suspended CNTs in the chamber using an evaporator heated to 480--520~$^\circ$C under a vacuum of less than $10^{-4}$~Pa. A glass slide is also placed in the chamber to quantify the deposition thickness from the absorbance of the CuPc peak. Calibration is performed by measuring the actual thickness with a surface profiler for two CuPc films with thicknesses of 80 and 167~nm. Three samples with different deposition amounts are prepared by changing the evaporation time, yielding nominal thicknesses on the substrate of 7, 16, and 26~nm.

\paragraph*{Suspended CNT field-effect transistor fabrication.}
The structures are fabricated on p-type Si substrates (resistivity $\sim 15 \pm 5$~m$\Omega\cdot$cm) coated with a 100~nm oxide layer~\cite{Yoshida:2016}. Trenches (500~nm deep, 0.4--1.6~$\mu$m wide) are formed by EBL and dry etching. The substrates are then oxidized at 900~$^\circ$C for 1~hour, creating an additional 20~nm of SiO$_2$ in the trenches. A second EBL step defines electrode patterns, followed by electron-beam evaporation of Ti (2~nm)/Pt (20~nm). After lift-off, a third EBL step patterns catalyst regions near the trenches, and CNTs are subsequently grown to span the trenches. The resulting suspended field-effect transistors enable efficient carrier-density modulation in the suspended CNT region, as confirmed by electrical characteristics and PL spectroscopy~\cite{Yoshida:2016,Yasukochi:2011}.

\paragraph*{Photoluminescence measurements.}
All PL measurements are performed at room temperature in a custom-built confocal microscopy setup purged with dry nitrogen gas. A Ti:sapphire laser, tunable in wavelength and operated in continuous-wave mode, serves as the excitation source. The laser power is controlled by neutral-density filters, and the beam is focused onto the sample via an objective lens (numerical aperture $\mathrm{NA}=0.65$, working distance $=4.5$~mm), producing a 1/e$^2$ beam diameter of $\sim1.16$~$\mu$m. A confocal pinhole sets the collection spot size to $\sim5.4$~$\mu$m in diameter. Emitted PL is collected by the same objective, dispersed by a spectrometer with a 150~lines/mm grating (dispersion $\sim0.52$~nm/pixel at 1340~nm), and detected with a liquid-nitrogen-cooled InGaAs diode array (1024~pixels). The reflected beam from the sample traces back the same path and is detected by a silicon photodiode. The excitation polarization angle is always adjusted to be parallel to the CNT axis unless specifically mentioned. For spatially resolved imaging in Fig.\,2, a three-dimensional motorized stage scans the heterostructure to generate PL excitation images. Because the collection spot exceeds the excitation spot, photons emitted outside the immediate laser focus are also collected, and the laser spot size ultimately sets the spatial resolution. Consequently, these images reflect the excitation efficiency profiles for the PL and help identify non-local excitation phenomena. For the CNT/CuPc samples in Fig.\,3 and gated CNT/WSe$_2$ samples in Fig.\,4, we use a similar setup with an objective lens of $\mathrm{NA}=0.8$ (working distance $=3.4$~mm), producing a 1/e$^2$ beam diameter of $\sim0.94$~$\mu$m. Since numerical aperture for the PL measurement system of CNT/CuPc samples is different from the other two structures in Fig.~3i, emission collection efficiency is also considered in $T_{\mathrm{CNT}}$ efficiency and $E_{\mathrm{11}}$ efficiency evaluation (Supplementary Note 6).

\begin{acknowledgments}
Parts of this study are supported by JSPS (KAKENHI JP25K00056, JP25K17919, JP22F22350, JP24H01202, JP22K14624, JP22K14623, JP24K17627, JP23K13622, JP21H05233, JP23H02052, JP22H05445, JP21K04826, JP22H04957, JP23H00262, JP25K21704), JST (CREST JPMJMI22708192, JPMJCR24A5, JPMJCR25A1, ASPIRE JPMJAP2310), and MEXT (ARIM JPMXP1222UT1135, WPI). The authors acknowledge the Advanced Manufacturing Support Team at RIKEN for technical assistance. The authors thank Dr. Masahiro Yoshida for the experiments on gated suspended CNTs, Dr. Shunsuke Tanaka for the experiments on CNT/CuPc hybrids, and Dr. Akihiro Ishii for the time-resolved PL measurement of the suspended CNT field-effect transistor sample.
\end{acknowledgments}

\section*{Author Contributions}
N.F. carried out sample preparation and performed measurements. U.E. prepared gated-heterostructure samples and performed measurements. Y.R.C. assisted in sample preparation. S.F. and D.Y. contributed to the time-resolved PL. C.F.F. assisted in data analysis. S.M., K.K., K.U., and K.N. provided Nb-doped WSe$_2$ crystals. T.T. and K.W. provided hBN crystals. Y.K.K. supervised the project. N.F. and Y.K.K. co-wrote the manuscript, with all authors providing input and comments on the manuscript.

\end{document}